\documentclass[rsi,aps,twocolumn,noshowpacs,noshowkeys,superscriptaddress]{revtex4-1}
\usepackage{amsfonts}
\usepackage{amsmath}
\usepackage{amssymb}
\usepackage{graphicx}

\usepackage{textcomp}%
\newcommand{\vv}[1]{\mathbf{#1}}

\usepackage{color}

\begin{document}

\def\neel{Institut N\'{e}el, Universit\'{e} Grenoble Alpes - CNRS:UPR2940, 38042 Grenoble, France}
\def\lkb{Laboratoire Kastler Brossel - UMR, Paris, France}
\author{Francesco Fogliano$^{\star}$}
\affiliation{\neel}
\thanks{equal contributions}
\author{Benjamin Besga$^{\star}$}
\affiliation{\neel}
\thanks{equal contributions}
\author{Antoine Reigue$^{\star}$}
\affiliation{\neel}
\thanks{equal contributions}
\author{Philip Heringlake}
\affiliation{\neel}
\author{Laure Mercier de L\'{e}pinay}
\affiliation{\neel}
\author{Cyril Vaneph}
\affiliation{\lkb}
\author{Jakob Reichel}
\affiliation{\lkb}
\author{Benjamin Pigeau}
\affiliation{\neel}
\author{Olivier Arcizet}
\affiliation{\neel}
\email{olivier.arcizet@neel.cnrs.fr}

\title{Cavity nano-optomechanics in the ultrastrong coupling regime with ultrasensitive force sensors}

\begin{abstract}
In a canonical optomechanical system, mechanical vibrations are dynamically encoded on an optical probe field which reciprocally exerts a backaction force. Due to the weak single photon coupling strength achieved with macroscopic oscillators, most of existing experiments were conducted with large photon numbers to achieve sizeable effects, thereby causing a dilution of the original optomechanical non-linearity. Here, we investigate the optomechanical interaction of an ultrasensitive suspended nanowire inserted in a fiber-based microcavity mode. This implementation allows to enter far into the hitherto unexplored ultrastrong optomechanical coupling regime, where one single intracavity photon can displace the oscillator by more than its zero point fluctuations.
To fully characterize our system, we implement nanowire-based scanning probe measurements to map the vectorial optomechanical coupling strength, but also to reveal the intracavity optomechanical force field experienced by the nanowire.
This work establishes that the single photon cavity optomechanics regime is within experimental reach.
\end{abstract}

\maketitle

{\it Introduction--}
The field of optomechanics has gone through many impressive developments over the last decades \cite{Aspelmeyer2014}. The coupling between a probe light field and a mechanical degree of freedom, an oscillator, possibly assisted by a high finesse cavity was early proposed as an ideal platform to explore the quantum limits of ultrasensitive measurements, where the quantum fluctuations of the light are the dominant source of measurement noise \cite{Braginsky1995,Caves1980,Jaekel2007,Fabre1994}. The measurement backaction was also employed to manipulate the oscillator state through optical forces and dynamical backaction, leading to optomechanical correlations between both components of the system. In this framework,  ground state cooling, mechanical detection of radiation pressure quantum noise, advanced correlation between light and mechanical states  or optomechanical squeezing were reported \cite{chan2011,Teufel2011,Purdy2013,Peterson2016,Kampel2017,Rossi2018,Weis2010,Verhagen2012,Palomaki2013,Lecocq2015b, Riedinger2016, Sudhir2017,Sudhir2017a,Pirkkalainen2015}.\\
All those impressive results were obtained in the linear regime of cavity optomechanics, making use of large photon numbers, where the interaction Hamiltonian is linearized around an operating setpoint. However, the optomechanical interaction possesses an intrinsic non-linearity at the single excitation level, which has for the moment remained far from experimental reach due to the weak single photon coupling strength achieved with macroscopic oscillators.\\
This regime is achieved when a single photon in the cavity shifts the static rest position of the mechanical resonator by a quantity $\delta x^{(1)}$  which is larger than its zero point fluctuations $\delta x_{\rm zpf}$. A very strong optomechanical interaction is indeed needed to fulfil this condition since it requires $g_{0} / \Omega_{\rm m} > 1 $ where $g_{0}$ is the single photon optomechanical coupling and $\Omega_{\rm m}$ the resonant pulsation of the mechanical oscillator. Operating in the ultra-strong coupling regime is thus an experimental challenge pursued by the community to open the box of single-photon cavity optomechanics.\\
\begin{figure}[b!]
\includegraphics[width=0.99\columnwidth]{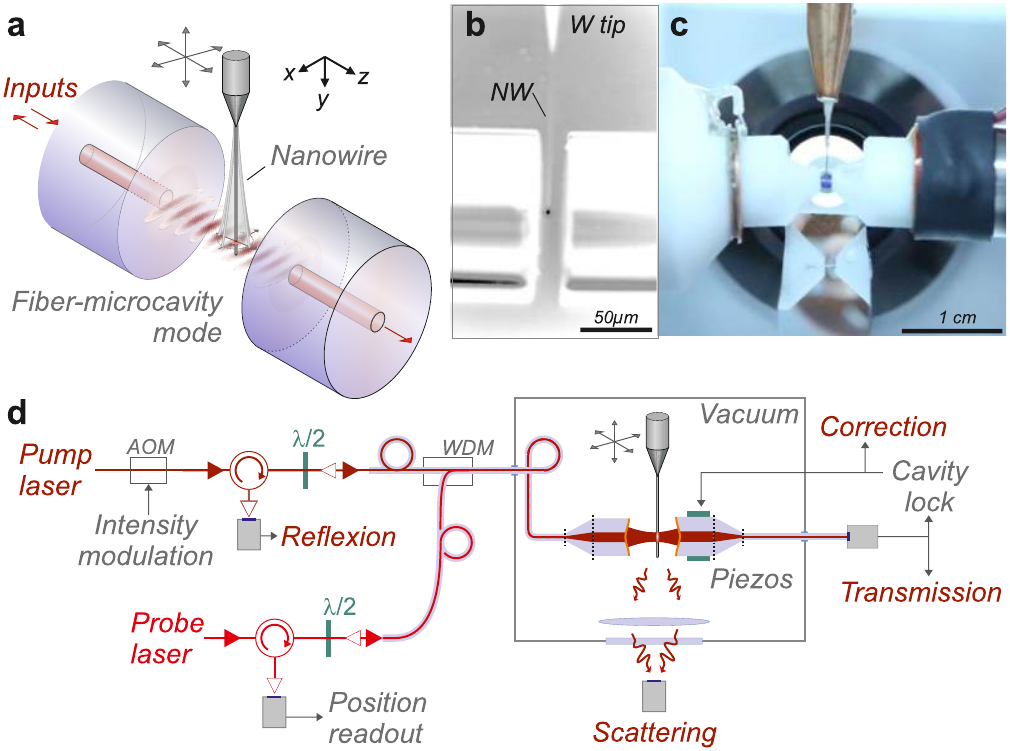}
\caption{
\textbf{Experimental setup.} {\bf a} the vibrating extremity of a silicon carbide nanowire is piezo-positioned in the optical mode volume of a high-finesse fiber microcavity to produce a large parametric coupling between the cavity optical mode and the oscillator position. {\bf b,c} A lateral  objective serves to align the experiment and collect part of the side-scattered light (dark spot in {\bf b}).
{\bf d} Sketch of the experiment. AOM: Acousto optic modulator; WDM wavelength dividing module.\\}
\label{Fig1}
\end{figure}
In this article, we report on a novel implementation, that operates far inside the ultrastrong coupling regime ($g_{0} / \Omega_{\rm m} > 100 $). To do so we make use of an ultrasensitive force sensor, a suspended silicon carbide nanowire with sub-wavelength sized diameter whose vibrating extremity is inserted in the optical mode of a high finesse fiber microcavity (see Fig.\,1a). This "nanowire in the middle" configuration decouples mechanics from optics \cite{Thomson2008, Favero2009, Anetsberger2009, Sankey2010, Purdy2013, Stapfner2013, Favero2014, Kiesel2013,Jockel2015,Wilson2015, Peterson2016, Sudhir2017, Sudhir2017a, Rossi2018} and makes use of ultra-sensitive nanowires as mechanical resonators \cite{Arcizet2011,Gloppe2014,Pigeau2015,Mercier2016,Rossi2016,Mercier2018} coupled to very short fiber Fabry-P\'erot microcavities \cite{Colombe2007, Hunger2010a, Mader2015}.
This approach first allows to explore the optical properties of the cavity mode using scanning probe imaging techniques. We measure the  dependence on the nanowire position of the frequency and linewidth of the cavity mode. This allows to spatially map the parametric optomechanical coupling strength, which acquires a vectorial character due to the ability of the nanowire to vibrate along both transverse directions in the cavity mode. Furthermore, we investigate the other facet of the optomechanical interaction  and realize for the first time a full mapping of the optomechanical force experienced by the nanowire. Those force measurements are realized using pump-probe techniques \cite{Gloppe2014} and provide a novel analytic tool of the intracavity field, complementary to previous optical measurements.
We conclude by discussing new observations that would become accessible when working with cryogenically cooled nanowires in the emerging field of single photon optomechanics.
\\
Operating in such a regime is extremely promising. Among other outcomes, it gives access to optomechanically based QND measurements of light intensity at ultra low photon fluxes or permits engineering optomechanical systems presenting non-linearities at the single intracavity photon level. This has stimulated a lot of theoretical work \cite{Bose1997,Mancini1997,Ludwig2008,Nunnenkamp2011,Nunnenkamp2012,Nation2013,Rimberg2014} also motivated by the expected deviation from the commonly employed mean-field approximations. For the moment only atom-based optomechanical experiments \cite{Murch2008} could approach such a regime, while recent developments with highly deformable photonic crystal \cite{Leijssen2015,Leijssen2017} or trampoline \cite{Reinhardt2016} resonators represent interesting platforms in that perspective.\\

{\it Formalization--}
The optomechanical interaction between an optical cavity mode and a  single mechanical vibration mode (pulsations $\omega_0 | \Omega_{\rm m}$; ladder operators $\hat a, \hat a^\dagger\, \| \hat b, \hat b^\dagger$)  is canonically described by the coupling Hamiltonian $H_{\rm int}= \hbar g_0 \hat a^\dagger \hat a (\hat b+\hat b^\dagger)$, which formalizes the parametric dependence of the optical cavity resonance pulsation $\omega_0(\hat x)$ on the oscillator position $\hat x=  \delta x_{\rm zpf} (\hat b+\hat b^\dagger)$, where  $\delta x_{\rm zpf}=\sqrt{\hbar/2M_{\rm eff}\Omega_{\rm m}}$ is the spatial spreading of the oscillator zero point fluctuations (effective mass $M_{\rm eff}$). The single photon parametric coupling strength $g_0=\partial_x \omega_0 \, \, \delta x_{\rm zpf}$ is maximium for oscillators featuring large zero-point fluctuations- motivating the shift towards ultralight nanowires- interacting with small mode volume cavity modes. A single photonic excitation ($\langle\hat a^\dagger \hat a \rangle = 1$) generates an optical force of $ F^{(1)} = - \hbar g_0 /\delta x_{\rm zpf} $ which causes a static displacement of $\delta x^{(1)}= F^{(1)} / M \Omega_{\rm m}^2 = 2 g_0/\Omega_{\rm m} \delta x_{\rm zpf}$. As such, a single intracavity photon can be expected to have an appreciable impact on the oscillator state when $\delta x^{(1)}> \delta x_{\rm zpf}$. This inequality is fulfilled if $ g_0>\Omega_{\rm m} / 2$, which defines the ultrastrong coupling regime of the parametric interaction.
Up to now the achieved single photon coupling strength $g_0$, remained too faint to have any impact on the oscillator state and all experiments were thus realized with large photon numbers to enhance the optomechanical interaction. In that situation, the system dynamics can be described by the linearized Hamiltonian: $\hbar g_0\bar\alpha   \left(\hat a+\hat a^\dagger\right) (\hat b+\hat b^\dagger)$, where the effective coupling strength $g_0\bar\alpha$ is formally enhanced by the mean intracavity field $\bar\alpha$. However, by doing so the fundamental optomechanical non-linearity- the optomechanically induced optical dephasing is proportional to the intracavity intensity like in the Kerr effect- gets diluted, and the regime of single photon optomechanics has remained for the moment out of experimental reach.


{\it The experiment--}
The fiber microcavity is made of 2 single mode fibers with concavely laser machined extremities covered with high reflectivity dielectric coatings ($\approx 28\,\rm \mu m$ radii of curvature) \cite{Hunger2012}. The optical finesse can be adjusted from 400 to 40\, 000 by tuning the laser wavelength from 760\,nm to 820\,nm. The cavity geometry (fiber angles, lateral position and coarse axial length) is adjusted using motorized supports, while the cavity length can be finely tuned using a set of piezo elements for fast and slow actuation (see SI).
To minimize the optical mode volume, we operate with cavity lengths down to $10\,\rm\mu m$ -which still permits to safely insert the nanowire between the fibers (see Fig.\,1ab)- producing an optical waist of $w_0 \approx 1.8\,\rm\mu m$.
Our nanomechanical probes are silicon carbide nanowires\cite{Gloppe2014,Pigeau2015,Mercier2016,Mercier2018}, mounted on sharp metallic tips, whose vibrating extremity can be finely positioned within the cavity mode using an XYZ piezo stage. The nanowires employed oscillate in the 10-100\,kHz range, with effective masses around 1\,pg,  and sub-wavelength sized diameters (100-200\,nm). The experiment is conducted in static vacuum (below $10^{-2}\,\rm mbar$), where quality factors around $5000$ are achieved, in a suspended vacuum chamber to avoid vibrations and thermally insulated to improve its long term stability.
The micro-cavity is pumped by a low-noise tunable infrared Ti:saphire laser. We record the transmitted and reflected signals, but also the side-scattered
photons, collected through a laterally positioned microscope objective (see Fig.\,1cd). A digital feedback loop acting on the cavity piezos can be activated to lock the cavity length on resonance.\\

\begin{figure*}[t!]
\begin{center}
\includegraphics[width=0.99 \linewidth]{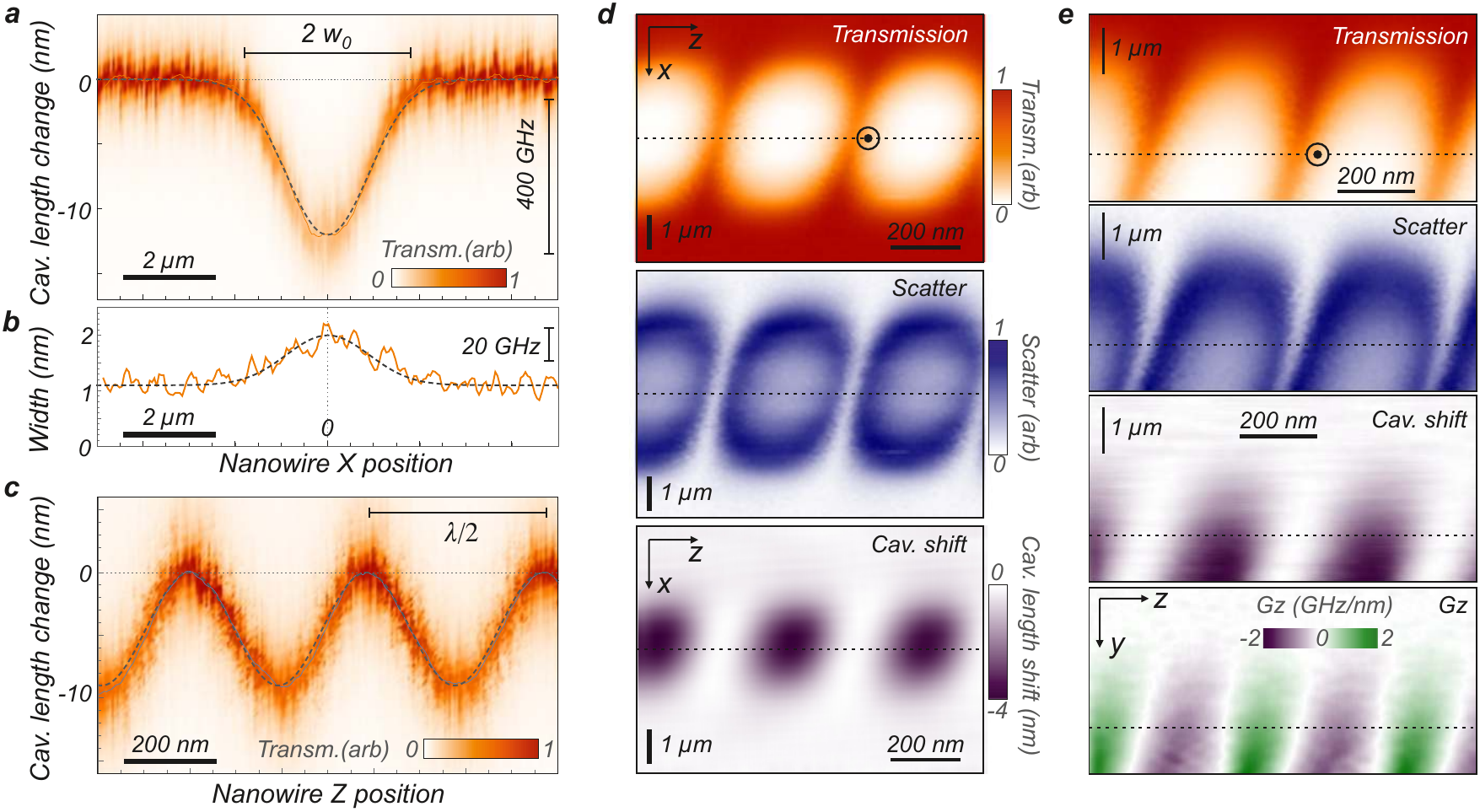}
\caption{
\textbf{Nanowire based scanning probe exploration of the intracavity field.}
{\bf a} Transmission stack of the cavity measured while scanning its total length with the slow piezo, in the vicinity of a $\rm TM_{00}$ mode and moving the nanowire across the optical mode (x axis). The dashed line is a gaussian fit of the measured resonance length shifts ($1.8\,\rm \mu m$ waist). The change in linewidth is reported in b. {\bf c} Similar measurement realized while scanning the nanowire along the optical axis, revealing the standing wave structure of the cavity mode. {\bf d} Transmission, scatter maps measured when scanning the nanowire in the XZ plane while locking the cavity on resonance. The cavity length shift map is obtained via the correction of the feedback loop, after substraction of the slow experimental drift. {\bf e} Similar measurements obtained when scanning the nanowire in the YZ plane (X=0), the optical axis is marked with a dashed line. The parametric coupling strength  $G_z$ is obtained by spatial derivation of the cavity shift map along $\vv{e_z}$.
}
\label{Fig2}
\end{center}
\end{figure*}


{\it Optomechanical coupling strength--}
We first determine the parametric optomechanical coupling strength by measuring the dependence of the cavity resonance $\omega_0(\vv{r})$ on the nanowire position $\vv{r}$. To do so we scan the cavity length with the slow piezo stack around a mean value of $ L = 12\,\rm \mu m$  and insert the nanowire in the optical mode volume. Fig.\,2a represents the cavity transmission recorded when pumped at 767\,nm in the vicinity of a $\rm TM_{00}$ mode while scanning a 130\,nm-thick-nanowire across the optical mode volume (along x axis). This causes a cavity frequency shift, a modification of its linewidth and a change of the resonant transmission level. The nanowire diameter (and effective optical cross-section \cite{Bohren1983}) being much smaller than the optical wavelength,  one recognizes here the lateral gaussian shape of the $\rm TM_{00}$ cavity mode, with a fitted optical waist of $1.8\,\rm\mu m$. When fully inserted (below the optical axis), the nanowire parametrically shifts the resonant cavity length by $\Delta L= -12\,\rm nm$, corresponding to an equivalent frequency shift of $\Delta\omega_0 = -\omega_0\,  \Delta L/L= 2\pi\times 400\, \rm GHz$. It corresponds to a $ 10^{-3} $ relative cavity shift, in agreement with the relative increase of the optical mode volume (see SI). Similar measurements can be realized while scanning the nanowire along the optical axis ($\vv{e_z}$), see Fig.\,2c, revealing the standing wave structure of the cavity mode, with $\lambda/2$ periodicity. When the nanowire is positioned on a node of the optical mode, the cavity field remains unperturbed.\\
Those measurements allow to spatially map the parametric coupling strength. Since the nanowire can move identically along both transverse  (xz) directions, it is necessary to adopt a vectorial coupling strength $\vv{G}\equiv \left.\boldsymbol{\nabla} \omega_0\right|_{\vv{r_0}}$, which has to be evaluated at the nanowire rest position $\vv{r_0}$. The maximum slopes observed amount to $G_x/2\pi\approx 0.3\,\rm GHz/nm$ and $G_z/2\pi \approx 3\,\rm GHz/nm$. Those measurements were realized with a nanowire featuring zero-point-fluctuations spreading over $\delta r_{\rm zpf}=\sqrt{\hbar/2M_{\rm eff}\Omega_{\rm m}}\approx 0.4\,\rm pm$, so that the single photon vectorial coupling strength ($\vv{g_0} \equiv \vv{G}\ \delta r_{\rm zpf} $ ) is as large as $g_0^z/2\pi= 1.2\,\rm MHz$.
This value, already rather large compared to other implementations \cite{Aspelmeyer2014}, is significantly larger than the nanowire fundamental frequency ($\Omega_{\rm m}/2\pi=50\,\rm kHz$ here), thus largely entering the ultrastrong coupling regime of the parametric interaction ($g_0/\Omega_{\rm m}= 25$). The maximum linewidth broadening caused by the nanowire thermal noise \cite{Leijssen2017}, spreading over $\Delta r^{\rm th} = \sqrt{k_B T/M_{\rm eff}\Omega_{\rm m}^2}\approx 7\,\rm nm $ amounts to $ G_z\Delta r^{\rm th} \approx 2\pi\times 23\,\rm GHz$ at 300K which becomes comparable to the broadened optical resonance (see Fig.2b) but remains small compared to the observed parametric shift (400\, GHz).


{\it Nanowire based characterization of the intracavity field--}
In view of its sub-wavelength sized diameter, the nanowire can be used to map the intracavity mode structure. Due to the finite laser mode-hope-free tunability ($\approx 30\,\rm GHz$), we chose to lock the cavity on the laser wavelength and thus compensate with the cavity piezos the nanowire induced optical resonance shifts. The error signal is synthesized from the cavity transmission signal using a 250\,kHz lock-in. We use a dual feedback loop acting on the fast and slow piezo elements with a bandwidth intentionally restricted to a few kHz in order not to compensate for resonant mechanical vibrations of the nanowire. Fig.\,2d(e), represents the experimental transmission, scatter (laterally collected through the objective) and correction signals recorded when scanning the nanowire in the horizontal XZ (vertical YZ) plane. This permit a direct visualization of the intracavity standing wave, as revealed in the transmission maps. The intracavity nodes (anti-nodes) appear as regions of large (low) transmission.\\
The scatter maps reflect a different behavior. Inserting progressively the nanowire in the cavity mode first causes an increase of the scattered light, as expected. However, this is followed by a reduction (white central areas), which simply originates from the reduction of the cavity finesse due to the increased loss rate. This transition from a nanowire scattering rate smaller or larger than the bare cavity loss rate is responsible for the ring shapes observed in Fig.\,2d. The same considerations also explain the signatures observed on the vertical maps, where the dispersive and dissipative optomechanical coupling rates can be tuned by adjusting the vertical insertion of the nanowire extremity in the cavity mode. As long as the nanowire stays out of a scatter ring, the cavity finesse remains unchanged at first order. Interestingly, positions on each side (along $\vv{e_z}$) of the optical nodes where the cavity finesse remains almost unchanged also feature large parametric coupling strength, as shown in Fig.\,3e (panel $G_z$) and are where the largest optomechanical backaction can be expected.\\
Those multi-channels scanning probe maps are influenced by several parameters, such as the nanowire diameter, laser wavelength and polarization, the cavity geometry (curvature, length, alignment) and finesse. They can all be tested quantitatively through this approach, which thus presents a significant analytical interest to investigate the intracavity field in confined microcavities.\\
\begin{figure*}[t!]
\begin{center}
\includegraphics[width=0.99 \linewidth]{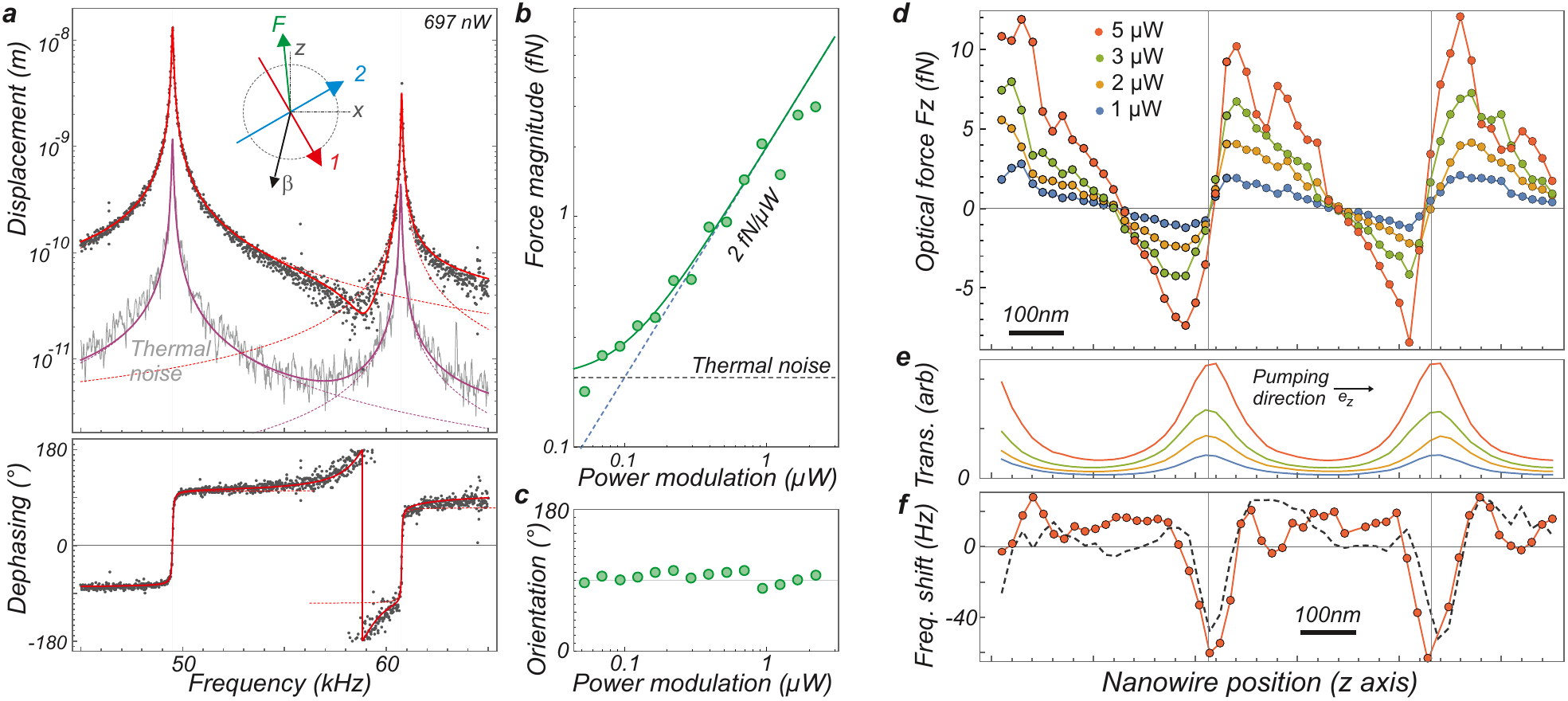}
\caption{
\textbf{Nanomechanical measurement of the intracavity optical force.} {\bf a}  Thermal noise (gray line, 3 Hz resolution bandwidth) and response measurements (dots) measured with the red probe laser  (XXX$\,\rm \mu W$) providing a projective measurement of the nanowire vibrations along the $\vv{e_\beta}$ axis (inset), for a cavity locked at resonance at position $\odot$ in Fig.\,2de. The response measurements are realized with a vectorial network analyzer, whose output channel serves to intensity modulate the infrared cavity pump laser. The phase difference is measured between the recorded displacement and the generated intensity modulation. Fitting of the data (see text) allows to determine the eigenmode orientations  $\vv{e_{1,2}}$, and the instantaneous optical force vector. {\bf b,c} Evolution of the magnitude and orientation of the optical force vector for increasing modulation depth ($P_0=3\, \rm\mu W$), illustrating the linearity of the system response. {\bf d} Dependence of the instantaneous force $F_z$ with the position along the optical axis for increasing input powers. The nodes of the electromagnetic modes are indicated as dashed lines and correspond to maxima in the transmissions plots (panel {\bf e}).{\bf f} Measured relative frequency shifts (dashed lines) obtained for $5\,\rm \mu W$ compared to the shifts expected from the measured optical force field gradients $\frac{-1}{2\Omega_{\rm m} M_{\rm eff}}\partial_z F_z $ (red dots).
 }
\label{Fig3}
\end{center}
\end{figure*}


{\it Nano-optomechanical investigation of the intracavity force--}
The above measurements allowed to investigate how the oscillator perturbs the intracavity field, which represents one facet of the optomechanical interaction. To fully quantify an optomechanical system it is essential to characterize the second facet of the interaction and measure the action of the intracavity light field on the oscillator. This dual characterization is essential, in particular for the "oscillator in the middle" approaches when the nature and excitation level of the optical mode involved in the interaction Hamiltonian (described by the $\hat a,\, \hat a ^\dagger$ operators) strongly depends on the oscillator position (optical mode spatial profile, oscillator induced losses...) and on the pumping conditions. We  measure the intracavity force field using a pump-probe scheme \cite{Gloppe2014} realized by modulating the intracavity intensity at frequencies close to the mechanical resonances, while recording the laser driven nanowire trajectories with a separate probe laser.\\
The visible probe laser (few $\rm\mu W$ at 633\,nm falling outside of the microcavity coating reflection window) is co-injected in the cavity fibers. Its reflection $P_R(\vv{r})$ which is recorded on a separate photodiode (see Fig\,1d and SI) presents an interference pattern (between reflections on the nanowire and input fiber extremity) that strongly depends on the nanowire position $\vv{r}=\vv{r_0}+\boldsymbol{\delta}\vv{r}$. The vibrations of its extremity $\vv{\delta r}(t)$ are dynamically encoded as photocurrent fluctuations $\delta P_R(t)=  \boldsymbol{\delta}\vv{r}(t)\cdot \left. \boldsymbol{\nabla}P_R\right|_{\vv{r_0}}$ which permits to measure the nanowire vibrations $\delta r_\beta (t)=\vv{\delta r} (t)\cdot \vv{e_\beta}$ projected along the local measurement vector $\vv{e_\beta}\equiv \boldsymbol{\nabla}P_R/|\nabla P_R|$  \cite{Gloppe2014,Mercier2016,Mercier2018}.
This allows probing the nanowire thermal noise (see Fig.\,3a grey curve) but also its response to an external force such as the one exerted by the intracavity light field.
\begin{figure}[t!]
\begin{center}
\includegraphics[width=0.99 \linewidth]{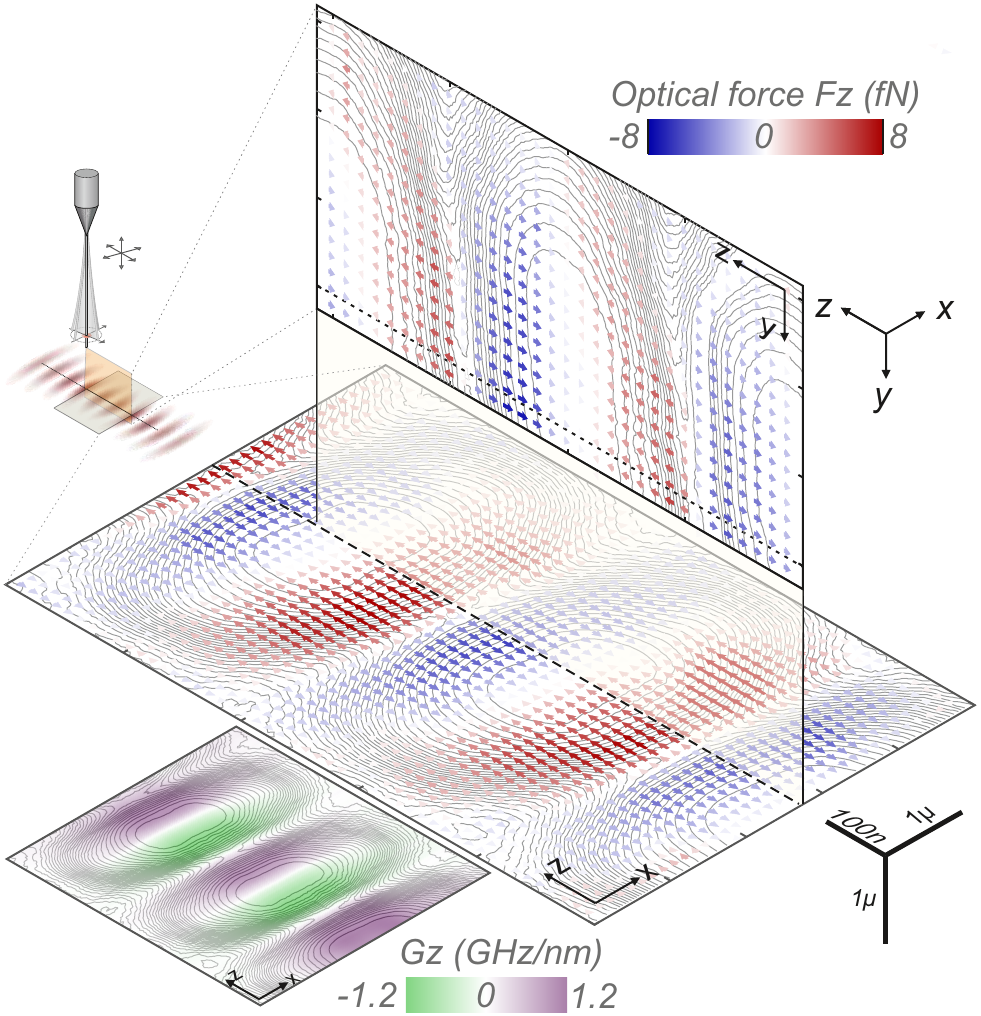}
\caption{
\textbf{Nano-optomechanical mapping of the intracavity force field.} Horizontal  (XZ) and vertical  (YZ) maps of the intracavity force field (anisotropic length scales), measured for an injected power of $3\,\rm \mu W$ at $767\,\rm nm$. The pump laser is injected along $\vv{e_z}$ and the dashed line represents the optical axis. The color code indicates the force component along the optical axis $F_z$. Inset: parametric coupling strength $G_z$ derived from the correction of the cavity lock. The transmission iso-values are superimposed as gray lines to help visualizing the intracavity field structure.}
\label{Fig4}
\end{center}
\end{figure}
During those measurements, the cavity is locked on the pump laser, whose intensity is partially time-modulated using an acousto-optic modulator: $P_0+\delta P \cos\Omega t$ with an amplitude $\delta P$ around a mean value $P_0$. The modulation frequency $\Omega/2\pi$ is swept across both eigenmode frequencies using a vectorial network analyser \cite{Gloppe2014} and we record the optically induced driven displacement $\delta r_\beta [\Omega]$ as shown in Fig.\,3a (black dots). The individual responses of each transverse mechanical mode are well distinguishable. The responses are well fitted in amplitude and phase with the complex expression:
$$\delta r_{\beta}[\Omega]= \vv{e_{\beta}}\cdot\sum_{i=1,2} \chi_i[\Omega](\vv{e_i}\cdot \vv{\delta F} )\vv{e_i} $$
where $\chi_i^{-1}[\Omega] \equiv M_{\rm eff}(\Omega_i^2-\Omega^2-i \Omega\Gamma)$, $(i=1,2)$ are the inverse mechanical susceptibilities of each eigenmodes.
The measurement vector $\vv{e_\beta}$ is independently determined using the standard calibration protocols \cite{Gloppe2014, Mercier2016}. The eigenmode orientations, frequencies, damping rates and effective masses are independently determined by fitting the Brownian motion spectra in absence of light modulation (see SI). Mode 1 is pointing at $+20^\circ$ from the optical axis $\vv{e_z}$. As such, the only fitting parameter is the modulated force complex vector $\vv{\delta F}$ caused by the intensity modulation $\delta P$ of the input light. Since the experiment is realized in the adiabatic cavity regime ($\Omega_{\rm m}\ll\kappa$), pure optical forces instantaneously follow (on mechanical timescales) the input light modulation, while photothermal forces will be in quadrature \cite{Gloppe2014}.
The amplitudes and orientations of the instantaneous optical force are reported in Fig.\,3bc for increasing modulation depth $\delta P$.
This permits to verify the linearity of the system with an approx. $2\,\rm fN/\mu W$ dependence of the optical force on the input optical power at a given position of the nanowire extremity in the optical mode (indicated by $\odot$ in Fig. 2de, slightly besides a node). As expected, the optical force is aligned with the optical axis ($\vv{e_z}$), as reported in the inset of Fig.\,3a. We note that the magnitude of the force is comparable to the $1/c$ force/power ratio of $3\,\rm fN/\mu W$ expected in case of a fully absorbed optical flux. This reflects the fact that the lateral cross section of the nanowire which remains small compared to the optical waist dimensions is well balanced by the cavity enhanced interaction, despite the very moderate finesse  $F\approx 200$ employed here to ensure a long term lock stability).\\
Identical response measurements were subsequently realized at various locations along the optical axis. The resulting instantaneous optical force is reported in Fig.\, 3d. It was measured with a fixed modulation depth $\delta P/P_0 \approx 0.7$ for increasing average input powers $P_0$ (from 1 to $5\,\rm \mu W$). The measured optical force is $\lambda/2$ periodic due to the standing wave periodicity. It has a repulsive character in the vicinity of the nodes of the intracavity field and on the contrary is found to be weakly attractive towards the antinodes of the intracavity field. This phenomenology is well explained by theoretical modeling (see SI) and can be understood with the following qualitative argument, which also holds for membrane in middle experiments \cite{Thomson2008}: regions of positive(resp. negative) dispersive shifts: $G_z > 0 (<0)$ are associated to situations where a larger amount of light is stored in the first(resp. second) sub-cavity, so that the total force is found positive(resp. negative). Also, the $\approx 30\%$ positive-negative asymmetry observed in the force extrema is connected to the laser pumping direction (along $\vv{e_z}$) and mirror asymmetries.\\
The intracavity force field exerted on the nanowire $\vv{F}(z)$ presents some large spatial variations of $\partial_z F_z \approx 1.5 \times 10^{-7}\,\rm N/m$ for $5\,\rm \mu W$ input power in the vicinity of the intracavity nodes. They add up to the intrinsic nanowire stiffness, $k=M_{\rm eff}\Omega_{\rm m}^2\approx 10^{-4}\,\rm N/m$, and are responsible for frequency softening at the nodes locations. The measured frequency shifts are shown in Fig.\,3f, and are in good agreement with the ones deduced from the gradient of force measurements. This validates the force measurement protocol exposed above. We point out that those force measurements were realized using a largely frequency split nanowire ($\approx 20\%$), in order not to experience eigenmode rotations, which would have otherwise required a vectorial 2D readout of the nanowire displacements \cite{Mercier2016,Mercier2018}.\\

Last, we pursued those measurements by mapping the intracavity 2D force field in 3D.  Here we employed a faster measurement scheme in order not to suffer from spatial drifts, which should remain smaller than 10\,nm over a few hours (corresponding to a $0.01^\circ$ temperature stability). Instead of a full response measurement (Fig.\,3a), we realized a simultaneous multi-frequency excitation drive at each position.  A triplet of 3 optical drive tones was employed per mechanical mode, separated by 50\,Hz, whose central frequency was locked onto the nanowire eigenfrequency with a soft peak tracking loop (see SI). This permits to simultaneously determine the local force, the mechanical frequencies and quality factors, while reducing the measurement time per point to 100\,ms, 100 times faster than full response measurements. The measurement vector $\vv{e_\beta}$ was also locally recorded in real-time using a 2 tones (80-85\,Hz) lock-in detection scheme (see SI).\\
The resulting XZ and YZ maps of the intracavity force field are shown in Fig.\,4, superimposed on the transmission maps encoded as iso-values to help localizing the intracavity field structure.  The color code shows the measured optical force projected along the optical axis ($ F_z $), shown as colored arrows in the maps. 
Those measurements were realized using 400x400 pixels, lasting over 4 hours, and the signal processing routine described in the SI employed signal averaging up to 10 neighboring pixels, which correspond to a spatial resolution of 25\,nm along the z axis.
One can recognize the anti-trapping / trapping structures in the vicinity of the nodes /antinodes, and the vanishing of the optical force when the nanowire is extracted out of the intracavity mode volume, both transversally (x) and vertically (y). The parametric coupling map ($G_z$) is derived by calculating the gradient of the cavity lock correction signal map (see Fig.\,2d, panel "cavity shift"), recorded during the same XZ measurement sequence. On each side of the antinodes, it is possible to identify locations of simultaneous large optical backaction and large parametric coupling strength. Those regions are clearly of great interest for future experiments in the ultrastrong coupling regime.\\
We note that operating with a quasi-frequency-degenerated nanowire and a full 2D readout scheme will allow future investigations of the possible non-reciprocal (non-conservative \cite{Mercier2018}) character of the intracavity force field and its impact on the nanowire dynamics. \\

{\it Conclusion and prospectives--}
We have demonstrated that the ultrastrong coupling regime can be largely achieved with the nanowire in the middle configuration. We realized a dual characterization of the optomechanical interaction, via nanowire mediated scanning probe measurements of the intracavity field and mapping of the intracavity optomechanical force field.
For consistency, all the experiments described in the manuscript were obtained with the same nanowire, but we note that larger coupling strength, up to $g_0^z/2\pi= 2.6\,\rm MHz\approx 100 \times \Omega_{\rm m}/2\pi $,  were reached with different nanowires, see SI. We note that the light-nanowire interaction can be further engineered and enhanced by exploiting higher order internal optical resonances \cite{Bohren1983}.
Operating with ultrasensitive kHz nanowires (see SI), such as the ones developed for operations at dilution temperatures \cite{Fogliano2019}, should permit to reach $\approx 7\,\rm MHz$ single photon coupling strength, while reducing their thermal noise spreading down to  $\approx 500$ times their zero point fluctuations when thermalized at 20\,mK (see SI). A single intracavity photon should then produce a static deformation of $\delta x^{(1)}\approx 8000  \delta x_{\rm zpf} $ largely detectable on top of the nanowire residual rms thermal noise, opening the road towards the direct detection of single photon recoil. Furthermore, this static deformation can in turn parametrically shift the cavity resonance, by a quantity $\delta \omega^{(1)}= g_0 \, \delta x^{(1)}/ \delta x_{\rm zpf} $. If it exceeds the cavity linewidth, $\delta \omega^{(1)}>\kappa$, the system can thus present a static optomechanical non-linearity at the single intracavity photon level. This exotic regime is largely achievable in our system since large single photon parametric cooperativity $\mathcal{C}^{(1)}\equiv 2 g_0^2/\Omega_{\rm m}\kappa \approx 10 $ can be achieved even with modest cavity linewidth (6 GHz). We also note that single photon cooperativities \cite{Aspelmeyer2014} as large as $g_0^2/\Gamma_{\rm m}\kappa \approx 4.2\times 10^5$ are within reach.  Those considerations are thus strong incentives for pushing those developments towards dilution temperatures, to explore for the first time with macroscopic oscillators the regime of single photon cavity optomechanics, where mean-field theories are not pertinent anymore.\\

{\it Acknowledgments--}
We warmly thank the PNEC group at ILM, T.\,Heldt for his assistance in the early phase of the experiment,   J.P.\, Poizat, G.\, Bachelier, J.\,Jarreau, C.\,Hoarau, E.\,Eyraud and D.\,Lepoittevin. This project is supported by the French National Research Agency  (JCJC-2016 CE09-2016-QCForce, FOCUS projects,  LANEF framework (ANR-10-LABX-51-01, project CryOptics and "Investissements d'avenir" program (ANR-15-IDEX-02, project CARTOF) and by the European Research Council under the EU's Horizon 2020 research and innovation programme, grant agreements No 671133 (EQUEMI project), 767579 (CARTOFF) and 820033 (AttoZepto).\\


\begin{thebibliography}{10}
\expandafter\ifx\csname url\endcsname\relax
  \def\url#1{\texttt{#1}}\fi
\expandafter\ifx\csname urlprefix\endcsname\relax\def\urlprefix{URL }\fi
\providecommand{\bibinfo}[2]{#2}
\providecommand{\eprint}[2][]{\url{#2}}

\bibitem{Aspelmeyer2014}
\bibinfo{author}{Aspelmeyer, M.}, \bibinfo{author}{Kippenberg, T.~J.} \&
  \bibinfo{author}{Marquardt, F.}
\newblock \bibinfo{title}{{Cavity optomechanics}}.
\newblock \emph{\bibinfo{journal}{Rev. Mod. Phys.}}
  \textbf{\bibinfo{volume}{86}} (\bibinfo{year}{2014}).

\bibitem{Braginsky1995}
\bibinfo{author}{Braginsky, V.~B.}, \bibinfo{author}{Khalili, F.~Y.} \&
  \bibinfo{author}{Thorne, K.~S.}
\newblock \emph{\bibinfo{title}{Quantum measurement}}
  (\bibinfo{publisher}{Cambridge University Press}, \bibinfo{year}{1995}).

\bibitem{Caves1980}
\bibinfo{author}{Caves, C.~M.}
\newblock \bibinfo{title}{Quantum-mechanical radiation-pressure fluctuations in
  an interferometer}.
\newblock \emph{\bibinfo{journal}{Phys. Rev. Lett.}}
  \textbf{\bibinfo{volume}{45}}, \bibinfo{pages}{75--79}
  (\bibinfo{year}{1980}).

\bibitem{Jaekel2007}
\bibinfo{author}{Jaekel, M.~T.} \& \bibinfo{author}{Reynaud, S.}
\newblock \bibinfo{title}{Quantum limits in interferometric measurements}.
\newblock \emph{\bibinfo{journal}{EPL (Europhysics Letters)}}
  \textbf{\bibinfo{volume}{13}}, \bibinfo{pages}{301} (\bibinfo{year}{2007}).

\bibitem{Fabre1994}
\bibinfo{author}{Fabre, C.} \emph{et~al.}
\newblock \bibinfo{title}{Quantum-noise reduction using a cavity with a movable
  mirror}.
\newblock \emph{\bibinfo{journal}{Phys. Rev. A}} \textbf{\bibinfo{volume}{49}},
  \bibinfo{pages}{1337--1343} (\bibinfo{year}{1994}).

\bibitem{chan2011}
\bibinfo{author}{Chan, J.} \emph{et~al.}
\newblock \bibinfo{title}{{Laser cooling of a nanomechanical oscillator into
  its quantum ground state}}.
\newblock \emph{\bibinfo{journal}{Nature}} \textbf{\bibinfo{volume}{478}},
  \bibinfo{pages}{89--92} (\bibinfo{year}{2011}).

\bibitem{Teufel2011}
\bibinfo{author}{Teufel, J.~D.} \emph{et~al.}
\newblock \bibinfo{title}{{Sideband cooling of micromechanical motion to the
  quantum ground state}}.
\newblock \emph{\bibinfo{journal}{Nature}} \textbf{\bibinfo{volume}{475}},
  \bibinfo{pages}{359--363} (\bibinfo{year}{2011}).

\bibitem{Purdy2013}
\bibinfo{author}{Purdy, T.~P.}, \bibinfo{author}{Peterson, R.~W.} \&
  \bibinfo{author}{Regal, C.~A.}
\newblock \bibinfo{title}{{Observation of Radiation Pressure Shot Noise on a
  Macroscopic Object}}.
\newblock \emph{\bibinfo{journal}{Science}} \textbf{\bibinfo{volume}{339}},
  \bibinfo{pages}{801--804} (\bibinfo{year}{2013}).
\newblock \eprint{1209.6334}.

\bibitem{Peterson2016}
\bibinfo{author}{Peterson, R.~W.} \emph{et~al.}
\newblock \bibinfo{title}{{Laser Cooling of a Micromechanical Membrane to the
  Quantum Backaction Limit}}.
\newblock \emph{\bibinfo{journal}{Physical Review Letters}}
  \textbf{\bibinfo{volume}{116}}, \bibinfo{pages}{063601}
  (\bibinfo{year}{2016}).
\newblock \eprint{1510.03911}.

\bibitem{Kampel2017}
\bibinfo{author}{Kampel, N.~S.} \emph{et~al.}
\newblock \bibinfo{title}{Improving broadband displacement detection with
  quantum correlations}.
\newblock \emph{\bibinfo{journal}{Phys. Rev. X}} \textbf{\bibinfo{volume}{7}},
  \bibinfo{pages}{021008} (\bibinfo{year}{2017}).

\bibitem{Rossi2018}
\bibinfo{author}{Rossi, M.}, \bibinfo{author}{Mason, D.},
  \bibinfo{author}{Chen, J.}, \bibinfo{author}{Tsaturyan, Y.} \&
  \bibinfo{author}{Schliesser, A.}
\newblock \bibinfo{title}{Measurement-based quantum control of mechanical
  motion}.
\newblock \emph{\bibinfo{journal}{Nature}} \textbf{\bibinfo{volume}{563}},
  \bibinfo{pages}{53--58} (\bibinfo{year}{2018}).

\bibitem{Weis2010}
\bibinfo{author}{Weis, S.} \emph{et~al.}
\newblock \bibinfo{title}{Optomechanically induced transparency}.
\newblock \emph{\bibinfo{journal}{Science}} \textbf{\bibinfo{volume}{330}},
  \bibinfo{pages}{1520--1523} (\bibinfo{year}{2010}).

\bibitem{Verhagen2012}
\bibinfo{author}{Verhagen, E.}, \bibinfo{author}{Deléglise, S.},
  \bibinfo{author}{Weis, S.}, \bibinfo{author}{Schliesser, A.} \&
  \bibinfo{author}{Kippenberg, T.}
\newblock \bibinfo{title}{Quantum-coherent coupling of a mechanical oscillator
  to an optical cavity mode}.
\newblock \emph{\bibinfo{journal}{Nature}} \textbf{\bibinfo{volume}{4824}},
  \bibinfo{pages}{63--67} (\bibinfo{year}{2012}).

\bibitem{Palomaki2013}
\bibinfo{author}{Palomaki, T.~a.}, \bibinfo{author}{Teufel, J.~D.},
  \bibinfo{author}{Simmonds, R.~W.} \& \bibinfo{author}{Lehnert, K.~W.}
\newblock \bibinfo{title}{{Entangling mechanical motion with microwave
  fields.}}
\newblock \emph{\bibinfo{journal}{Science (New York, N.Y.)}}
  \textbf{\bibinfo{volume}{342}}, \bibinfo{pages}{710--3}
  (\bibinfo{year}{2013}).
\newblock \eprint{arXiv:1011.1669v3}.

\bibitem{Lecocq2015b}
\bibinfo{author}{Lecocq, F.}, \bibinfo{author}{Clark, J.~B.},
  \bibinfo{author}{Simmonds, R.~W.}, \bibinfo{author}{Aumentado, J.} \&
  \bibinfo{author}{Teufel, J.~D.}
\newblock \bibinfo{title}{{Quantum Nondemolition Measurement of a Nonclassical
  State of a Massive Object}}.
\newblock \emph{\bibinfo{journal}{Physical Review X}}
  \textbf{\bibinfo{volume}{041037}}, \bibinfo{pages}{1--8}
  (\bibinfo{year}{2015}).

\bibitem{Riedinger2016}
\bibinfo{author}{Riedinger, R.} \emph{et~al.}
\newblock \bibinfo{title}{{Non-classical correlations between single photons
  and phonons from a mechanical oscillator}}.
\newblock \emph{\bibinfo{journal}{Nature}} \textbf{\bibinfo{volume}{530}},
  \bibinfo{pages}{313--316} (\bibinfo{year}{2016}).
\newblock \eprint{1512.05360}.

\bibitem{Sudhir2017}
\bibinfo{author}{Sudhir, V.} \emph{et~al.}
\newblock \bibinfo{title}{Appearance and disappearance of quantum correlations
  in measurement-based feedback control of a mechanical oscillator}.
\newblock \emph{\bibinfo{journal}{Phys. Rev. X}} \textbf{\bibinfo{volume}{7}},
  \bibinfo{pages}{011001} (\bibinfo{year}{2017}).

\bibitem{Sudhir2017a}
\bibinfo{author}{Sudhir, V.} \emph{et~al.}
\newblock \bibinfo{title}{Quantum correlations of light from a room-temperature
  mechanical oscillator}.
\newblock \emph{\bibinfo{journal}{Phys. Rev. X}} \textbf{\bibinfo{volume}{7}},
  \bibinfo{pages}{031055} (\bibinfo{year}{2017}).

\bibitem{Pirkkalainen2015}
\bibinfo{author}{Pirkkalainen, J.-M.}, \bibinfo{author}{Damsk{\"{a}}gg, E.},
  \bibinfo{author}{Brandt, M.}, \bibinfo{author}{Massel, F.} \&
  \bibinfo{author}{Sillanp{\"{a}}{\"{a}}, M.~A.}
\newblock \bibinfo{title}{{Squeezing of Quantum Noise of Motion in a
  Micromechanical Resonator}}.
\newblock \emph{\bibinfo{journal}{Physical Review Letters}}
  \textbf{\bibinfo{volume}{115}}, \bibinfo{pages}{1--5} (\bibinfo{year}{2015}).

\bibitem{Thomson2008}
\bibinfo{author}{Thompson, J.~D.} \emph{et~al.}
\newblock \bibinfo{title}{{Strong dispersive coupling of a high-finesse cavity
  to a micromechanical membrane}}.
\newblock \emph{\bibinfo{journal}{Nature}} \textbf{\bibinfo{volume}{452}}
  (\bibinfo{year}{2008}).

\bibitem{Favero2009}
\bibinfo{author}{Favero, I.} \emph{et~al.}
\newblock \bibinfo{title}{{Fluctuating nanomechanical system in a high finesse
  optical microcavity}}.
\newblock \emph{\bibinfo{journal}{Opt. Express}} \textbf{\bibinfo{volume}{17}},
  \bibinfo{pages}{12813--12820} (\bibinfo{year}{2009}).

\bibitem{Anetsberger2009}
\bibinfo{author}{Anetsberger, G.} \emph{et~al.}
\newblock \bibinfo{title}{Near-field cavity optomechanics with nanomechanical
  oscillators}.
\newblock \emph{\bibinfo{journal}{Nature Physics}}
  \textbf{\bibinfo{volume}{5}}, \bibinfo{pages}{909--914}
  (\bibinfo{year}{2009}).

\bibitem{Sankey2010}
\bibinfo{author}{Sankey, J.~C.}, \bibinfo{author}{Yang, C.},
  \bibinfo{author}{Zwickl, B.~M.}, \bibinfo{author}{Jayich, A.~M.} \&
  \bibinfo{author}{Harris, J. G.~E.}
\newblock \bibinfo{title}{{Strong and tunable nonlinear optomechanical coupling
  in a low-loss system}}.
\newblock \emph{\bibinfo{journal}{Nature Physics}}
  \textbf{\bibinfo{volume}{6}}, \bibinfo{pages}{707--712}
  (\bibinfo{year}{2010}).

\bibitem{Stapfner2013}
\bibinfo{author}{Stapfner, S.} \emph{et~al.}
\newblock \bibinfo{title}{{Cavity-enhanced optical detection of carbon nanotube
  Brownian motion}}.
\newblock \emph{\bibinfo{journal}{Applied Physics Letters}}
  \textbf{\bibinfo{volume}{102}}, \bibinfo{pages}{151910}
  (\bibinfo{year}{2013}).
\newblock \eprint{1211.1608}.

\bibitem{Favero2014}
\bibinfo{author}{Favero, I.}, \bibinfo{author}{Sankey, J.} \&
  \bibinfo{author}{Weig, E.~M.}
\newblock \emph{\bibinfo{title}{Mechanical Resonators in the Middle of an
  Optical Cavity}}, \bibinfo{pages}{83--119} (\bibinfo{publisher}{Springer
  Berlin Heidelberg}, \bibinfo{address}{Berlin, Heidelberg},
  \bibinfo{year}{2014}).

\bibitem{Kiesel2013}
\bibinfo{author}{Kiesel, N.} \emph{et~al.}
\newblock \bibinfo{title}{Cavity cooling of an optically levitated
  nanoparticle}.
\newblock \emph{\bibinfo{journal}{PNAS USA}} \textbf{\bibinfo{volume}{110}},
  \bibinfo{pages}{14180--14185} (\bibinfo{year}{2013}).

\bibitem{Jockel2015}
\bibinfo{author}{J{\"{o}}ckel, A.} \emph{et~al.}
\newblock \bibinfo{title}{{Sympathetic cooling of a membrane oscillator in a
  hybrid mechanical–atomic system}}.
\newblock \emph{\bibinfo{journal}{Nature Nanotechnology}}
  \textbf{\bibinfo{volume}{10}}, \bibinfo{pages}{55--59}
  (\bibinfo{year}{2015}).
\newblock \eprint{1407.6820}.

\bibitem{Wilson2015}
\bibinfo{author}{Wilson, D.~J.} \emph{et~al.}
\newblock \bibinfo{title}{Measurement-based control of a mechanical oscillator
  at its thermal decoherence rate}.
\newblock \emph{\bibinfo{journal}{Nature}} \textbf{\bibinfo{volume}{524}},
  \bibinfo{pages}{325--} (\bibinfo{year}{2015}).

\bibitem{Arcizet2011}
\bibinfo{author}{Arcizet, O.} \emph{et~al.}
\newblock \bibinfo{title}{{A single nitrogen-vacancy defect coupled to a
  nanomechanical oscillator}}.
\newblock \emph{\bibinfo{journal}{Nature Physics}}
  \textbf{\bibinfo{volume}{7}}, \bibinfo{pages}{879--883}
  (\bibinfo{year}{2011}).
\newblock \eprint{1112.1291}.

\bibitem{Gloppe2014}
\bibinfo{author}{Gloppe, A.} \emph{et~al.}
\newblock \bibinfo{title}{{Bidimensional nano-optomechanics and topological
  backaction in a non-conservative radiation force field}}.
\newblock \emph{\bibinfo{journal}{Nature Nanotechnology}}
  \textbf{\bibinfo{volume}{9}}, \bibinfo{pages}{920--926}
  (\bibinfo{year}{2014}).

\bibitem{Pigeau2015}
\bibinfo{author}{Pigeau, B.} \emph{et~al.}
\newblock \bibinfo{title}{Observation of a phononic mollow triplet in a
  multimode hybrid spin-nanomechanical system}.
\newblock \emph{\bibinfo{journal}{Nat Commun}} \textbf{\bibinfo{volume}{6}},
  \bibinfo{pages}{8603} (\bibinfo{year}{2015}).

\bibitem{Mercier2016}
\bibinfo{author}{{Mercier de L{\'{e}}pinay}, L.} \emph{et~al.}
\newblock \bibinfo{title}{{A universal and ultrasensitive vectorial
  nanomechanical sensor for imaging 2D force field}}.
\newblock \emph{\bibinfo{journal}{Nature Nanotechnology}}
  \textbf{\bibinfo{volume}{12}}, \bibinfo{pages}{156--162}
  (\bibinfo{year}{2016}).

\bibitem{Rossi2016}
\bibinfo{author}{{Rossi}, N.} \emph{et~al.}
\newblock \bibinfo{title}{Vectorial scanning force microscopy using a nanowire
  sensor}.
\newblock \emph{\bibinfo{journal}{Nature Nano}} \textbf{\bibinfo{volume}{12}},
  \bibinfo{pages}{150--155} (\bibinfo{year}{2017}).

\bibitem{Mercier2018}
\bibinfo{author}{{Mercier de L{\'{e}}pinay}, L.}, \bibinfo{author}{Pigeau, B.},
  \bibinfo{author}{Besga, B.} \& \bibinfo{author}{Arcizet, O.}
\newblock \bibinfo{title}{{Eigenmode orthogonality breaking and anomalous
  dynamics in multimode nano-optomechanical systems under non-reciprocal
  coupling}}.
\newblock \emph{\bibinfo{journal}{Nature Communications}}
  \textbf{\bibinfo{volume}{9}}, \bibinfo{pages}{1--10} (\bibinfo{year}{2018}).

\bibitem{Colombe2007}
\bibinfo{author}{Colombe, Y.} \emph{et~al.}
\newblock \bibinfo{title}{{Strong atom - field coupling for Bose - Einstein
  condensates in an optical cavity on a chip}}.
\newblock \emph{\bibinfo{journal}{Nature}} \textbf{\bibinfo{volume}{450}},
  \bibinfo{pages}{1--6} (\bibinfo{year}{2007}).

\bibitem{Hunger2010a}
\bibinfo{author}{Hunger, D.} \emph{et~al.}
\newblock \bibinfo{title}{{A fiber Fabry-Perot cavity with high finesse}}.
\newblock \emph{\bibinfo{journal}{New Journal of Physics}}
  \textbf{\bibinfo{volume}{12}} (\bibinfo{year}{2010}).
\newblock \eprint{1005.0067}.

\bibitem{Mader2015}
\bibinfo{author}{Mader, M.}, \bibinfo{author}{Reichel, J.},
  \bibinfo{author}{H{\"{a}}nsch, T.~W.} \& \bibinfo{author}{Hunger, D.}
\newblock \bibinfo{title}{{A scanning cavity microscope}}.
\newblock \emph{\bibinfo{journal}{Nature communications}}
  \textbf{\bibinfo{volume}{6}}, \bibinfo{pages}{1--7} (\bibinfo{year}{2015}).

\bibitem{Bose1997}
\bibinfo{author}{Bose, S.}, \bibinfo{author}{Jacobs, K.} \&
  \bibinfo{author}{Knight, P.~L.}
\newblock \bibinfo{title}{{Preparation of nonclassical states in cavities with
  a moving mirror}}.
\newblock \emph{\bibinfo{journal}{Physical review A}}
  \textbf{\bibinfo{volume}{56}} (\bibinfo{year}{1997}).

\bibitem{Mancini1997}
\bibinfo{author}{Mancini, S.}, \bibinfo{author}{Man'ko, V.} \&
  \bibinfo{author}{Tombesi, P.}
\newblock \bibinfo{title}{{Ponderomotive control of quantum macroscopic
  coherence}}.
\newblock \emph{\bibinfo{journal}{Physical Review A}}
  \textbf{\bibinfo{volume}{55}} (\bibinfo{year}{1997}).

\bibitem{Ludwig2008}
\bibinfo{author}{Ludwig, M.}, \bibinfo{author}{Kubala, B.} \&
  \bibinfo{author}{Marquardt, F.}
\newblock \bibinfo{title}{{The optomechanical instability in the quantum
  regime}}.
\newblock \emph{\bibinfo{journal}{New Journal of Physics}}
  \textbf{\bibinfo{volume}{10}} (\bibinfo{year}{2008}).

\bibitem{Nunnenkamp2011}
\bibinfo{author}{Nunnenkamp, A.}, \bibinfo{author}{Borkje, K.} \&
  \bibinfo{author}{Girvin, S.~M.}
\newblock \bibinfo{title}{{Single-photon optomechanics}}.
\newblock \emph{\bibinfo{journal}{Physical Review Letters}}
  \textbf{\bibinfo{volume}{107}}, \bibinfo{pages}{1--5} (\bibinfo{year}{2011}).
\newblock \eprint{1103.2788}.

\bibitem{Nunnenkamp2012}
\bibinfo{author}{Nunnenkamp, A.}, \bibinfo{author}{B{\o}rkje, K.} \&
  \bibinfo{author}{Girvin, S.~M.}
\newblock \bibinfo{title}{{Cooling in the single-photon strong-coupling regime
  of cavity optomechanics}}.
\newblock \emph{\bibinfo{journal}{Phys. Rev. A}} \textbf{\bibinfo{volume}{85}},
  \bibinfo{pages}{1--4} (\bibinfo{year}{2012}).

\bibitem{Nation2013}
\bibinfo{author}{Nation, P.~D.}
\newblock \bibinfo{title}{{Nonclassical mechanical states in an optomechanical
  micromaser analog}}.
\newblock \emph{\bibinfo{journal}{Physical Review A}}
  \textbf{\bibinfo{volume}{88}}, \bibinfo{pages}{1--7} (\bibinfo{year}{2013}).

\bibitem{Rimberg2014}
\bibinfo{author}{Rimberg, A.~J.}, \bibinfo{author}{Blencowe, M.~P.},
  \bibinfo{author}{Armour, A.~D.} \& \bibinfo{author}{Nation, P.~D.}
\newblock \bibinfo{title}{{A cavity-Cooper pair transistor scheme for
  investigating quantum optomechanics in the ultra-strong coupling regime}}.
\newblock \emph{\bibinfo{journal}{New Journal of Physics}}
  \textbf{\bibinfo{volume}{16}} (\bibinfo{year}{2014}).

\bibitem{Murch2008}
\bibinfo{author}{Murch, K.~W.}, \bibinfo{author}{Moore, K.~L.},
  \bibinfo{author}{Gupta, S.} \& \bibinfo{author}{Stamper-kurn, D.}
\newblock \bibinfo{title}{{Observation of quantum-measurement backaction with
  an ultracold atomic gas}}.
\newblock \emph{\bibinfo{journal}{Nature Physics}}
  \textbf{\bibinfo{volume}{4}}, \bibinfo{pages}{0--3} (\bibinfo{year}{2008}).

\bibitem{Leijssen2015}
\bibinfo{author}{Leijssen, R.} \& \bibinfo{author}{Verhagen, E.}
\newblock \bibinfo{title}{{Strong optomechanical interactions in a sliced
  photonic crystal nanobeam}}.
\newblock \emph{\bibinfo{journal}{Nature Publishing Group}}
  \bibinfo{pages}{1--10} (\bibinfo{year}{2015}).

\bibitem{Leijssen2017}
\bibinfo{author}{Leijssen, R.}, \bibinfo{author}{{La Gala}, G.~R.},
  \bibinfo{author}{Freisem, L.}, \bibinfo{author}{Muhonen, J.~T.} \&
  \bibinfo{author}{Verhagen, E.}
\newblock \bibinfo{title}{{Nonlinear cavity optomechanics with nanomechanical
  thermal fluctuation}}.
\newblock \emph{\bibinfo{journal}{Nature Communications}}
  \textbf{\bibinfo{volume}{8}}, \bibinfo{pages}{1--10} (\bibinfo{year}{2017}).

\bibitem{Reinhardt2016}
\bibinfo{author}{Reinhardt, C.}, \bibinfo{author}{M\"uller, T.},
  \bibinfo{author}{Bourassa, A.} \& \bibinfo{author}{Sankey, J.~C.}
\newblock \bibinfo{title}{Ultralow-noise sin trampoline resonators for sensing
  and optomechanics}.
\newblock \emph{\bibinfo{journal}{Phys. Rev. X}} \textbf{\bibinfo{volume}{6}},
  \bibinfo{pages}{021001} (\bibinfo{year}{2016}).

\bibitem{Hunger2012}
\bibinfo{author}{Hunger, D.}, \bibinfo{author}{Deutsch, C.},
  \bibinfo{author}{Barbour, R.~J.}, \bibinfo{author}{Warburton, R.~J.} \&
  \bibinfo{author}{Reichel, J.}
\newblock \bibinfo{title}{Laser micro-fabrication of concave, low-roughness
  features in silica}.
\newblock \emph{\bibinfo{journal}{AIP Advances}} \bibinfo{pages}{012119}
  (\bibinfo{year}{2012}).

\bibitem{Bohren1983}
\bibinfo{author}{Bohren, C.~F.} \& \bibinfo{author}{Huffman, D.}
\newblock \emph{\bibinfo{title}{Absorption and Scattering of Light by Small
  Particles}} (\bibinfo{publisher}{WileyVCH, Berlin}, \bibinfo{year}{1983}).

\bibitem{Fogliano2019}
\bibinfo{author}{Fogliano, F.~{\it et al}.}
\newblock \bibinfo{title}{Ultrasensitive nano-optomechanical force sensor at
  dilution temperatures}.
\newblock \emph{\bibinfo{journal}{in preparation}}  (\bibinfo{year}{2019}).

\end{thebibliography}
\end{document}